\documentclass{moriond}

\def\be{\begin{equation}}
\def\ee{\end{equation}}
\def\bea{\begin{eqnarray}}
\def\eea{\end{eqnarray}}

\newcommand{\Photo}{\includegraphics[height=35mm]{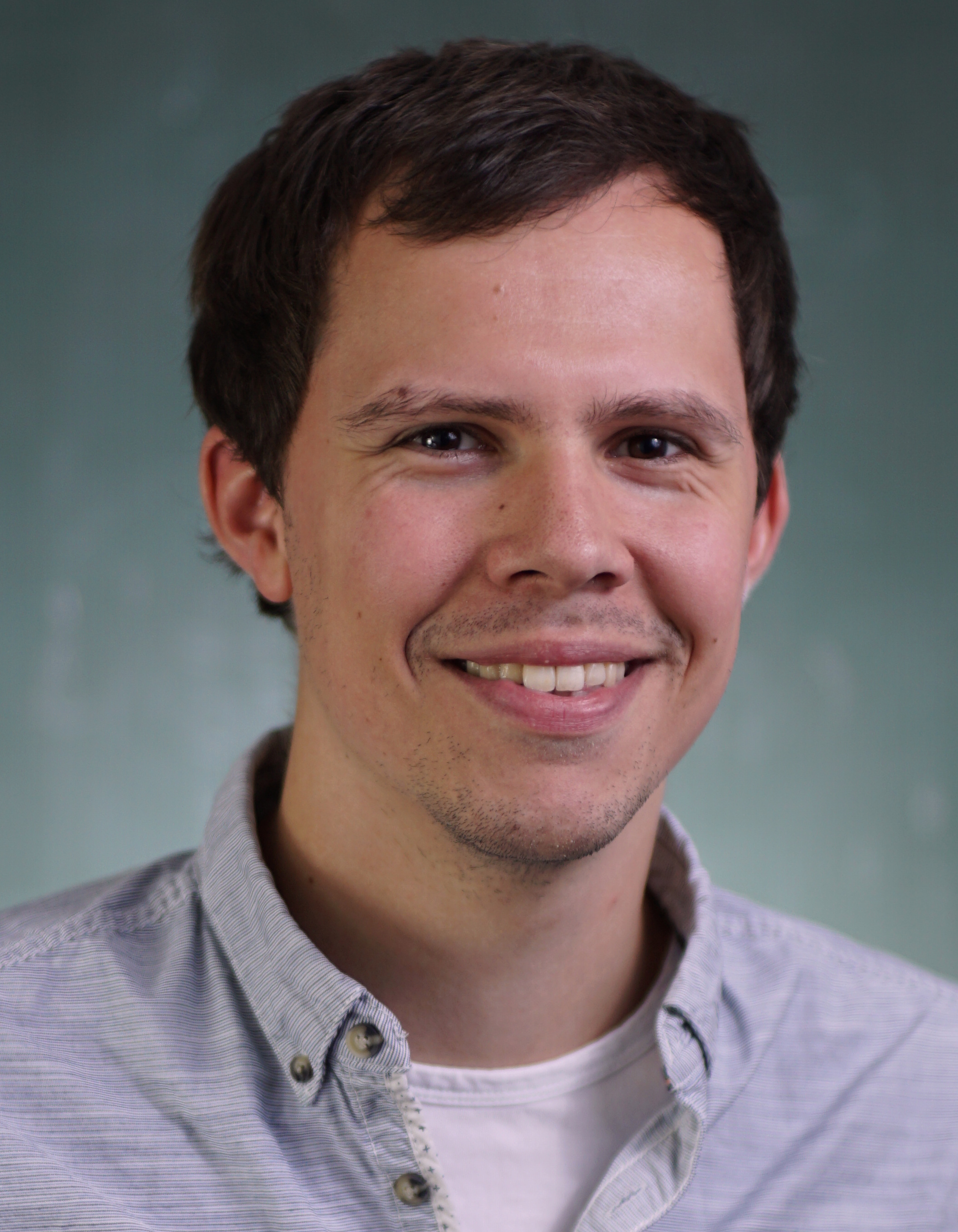}}

\begin{document}
\vspace*{4cm}
\title{Dark matter implications of the KATRIN neutrino mass experiment}
\author{Thede de Boer$^*$, Michael Klasen$^*$, Caroline Rodenbeck$^\dagger$, Sybrand Zeinstra$^{*}$\footnote{Speaker}}

\address{$^*$ Institut f\"ur Theoretische Physik, Westf\"alische Wilhelms-Universit\"at M\"unster, Wilhelm-Klemm-Stra\ss{}e 9, 48149 M\"unster, Germany \\
$^\dagger$ Institut f\"ur Kernphysik, Westf\"alische Wilhelms-Universit\"at M\"unster, Wilhelm-Klemm-Stra\ss{}e 9, 48149 M\"unster, Germany}

{\hfill MS-TP-21-15} \vspace{-10mm}
\maketitle\abstracts{
We studied the effects of the absolute neutrino mass scale in the scotogenic radiative seesaw model. From a scan over the parameter space of this model, a linear relation between the absolute neutrino mass and the dark sector-Higgs coupling $\lambda_5= 3.1\times10^{-9}\ m_{\nu_e}/$eV has been established. With the projected sensitivity of the KATRIN experiment nearing cosmologically favored values, a neutrino mass measurement would fix the value of $\lambda_5$. Subsequent correlations between the DM mass and the Yukawa coupling between DM and the SM leptons can probe the fermion DM parameter space, when lepton flavor violation constraints are also considered. The results are independent of the neutrino mass hierarchy and the CP phase. 
}
\section{Introduction}
\label{sec:1}
The exact nature of cold Dark Matter (DM) is one of the main open questions in particle physics, as the Standard Model (SM) does not contain a suitable candidate. The observation of atmospheric and solar neutrino oscillations has established mass difference between the three neutrino generations, implying that there exist at least two massive neutrinos. The absolute mass scale is still an open question. Recent measurements from the KATRIN experiment have established an upper bound on the electron neutrino mass of 1.1 eV \cite{Aker:2019uuj}, with the goal of lowering this bound to the value of 0.2 eV \cite{Drexlin:2013lha}, putting it on the same order of magnitude as the cosmological bounds on the sum of the SM neutrino masses $\sum_i m_{\nu_i} < 0.12$ eV, assuming normal hierarchy (NH) and the $\Lambda$CDM model \cite{Aghanim:2018eyx}. An inverted hierarchy (IH) is also still a possibility. 

In radiative seesaw models, neutrino masses are generated at one-loop level through interaction with a dark sector that allows for DM candidates. The scotogenic model, proposed by Ma, features a dark sector that includes one scalar doublet $\eta$, and three generations of fermionic singlets $N_i$ ($i=1,2,3$) which are required to generate three non-zero SM neutrino masses\cite{Ma:2006km}. 

In our paper\cite{deBoer:2020yyw} we show that a measurement of the absolute neutrino mass can provide stringent constraints on the scotogenic model by means of a parameter scan through its parameter space.
\section{The scotogenic model}
\label{sec:2}
The SM is extended by a dark sector with an inert scalar doublet $\eta$ and three generations
of fermion singlets $N_i$ \cite{Ma:2006km}. The additional terms in the Lagrangian are
\bea
 \mathcal{L} &=&-\frac{m_{N_i}}{2}N_iN_i+y_{i\alpha}(\eta^\dagger L_\alpha) N_i
+\mathrm{h.c.}- m_{\phi}^2\phi^\dag\phi-m_\eta^2\eta^\dag\eta -\!\frac{\lambda_1}{2}\left(\phi^\dag\phi\right)^2 \\
&-&\!\frac{\lambda_2}{2}\left(\eta^\dag\eta\right)^2
\!-\lambda_3\left(\phi^\dag\phi\right)
\left(\eta^\dag\eta\right)
-\lambda_4\left(\phi^\dag\eta\right)\left(\eta^\dag\phi\right)
-\frac{\lambda_5}{2}\left[\left(\phi^\dag\eta\right)^2
+\left(\eta^\dag\phi\right)^2\right] \nonumber
\eea
where the three generations of left-handed SM lepton doublets are denoted by $L_\alpha$ ($\alpha=1,2,3$). In this work we assume that the lightest fermion singlet $N_1$ with mass $m_{N_1}$ is the DM candidate. $\phi$ denotes the SM Higgs field, which fixes the parameters $m_\phi$ and $\lambda_1$. Its vacuum
expectation value (VEV) is $\langle\phi^0\rangle=246$ GeV$/\sqrt{2}$ \cite{Tanabashi:2018oca}. We ensure that the scalar potential is bounded from below, and impose the perturbativity constraint $|\lambda_{2,3,4,5}|<4\pi$. $\eta$ does not acquire a VEV, thus we set $\lambda_2=0.5$ without affecting the phenomenology. The components of $\eta$ result in three physical mass eigenstates
\bea
m_{\eta^+}^2= m_\eta^2+\lambda_3\langle\phi^0\rangle^2, \hspace{2cm}
m_{R,I}^2= m_{\eta}^2+\left(\lambda_3+\lambda_4\pm\lambda_5\right)\langle\phi^0\rangle^2
\eea
The scalars from the neutral component of the scalar doublet $\eta_R$ and $\eta_I$ have masses $m_R$, $m_I$ with $+\lambda_5$ and $-\lambda_5$ respectively that are very close due to the natural smallness of $\lambda_5$ \cite{Kubo:2006yx}. We scan over the range $10^{-12}< |\lambda_5|<10^{-8}$ following previous work \cite{Vicente:2014wga}. We sample in the range $m_\eta\in[0.1;10]$ TeV, in accordance with the LEP lower bound on charged particle masses \cite{Abbiendi:2003yd}, which we also use as sampling range for the sterile neutrino masses $m_{N_i}$.

\section{Experimental constraints}
\label{sec:3}
SM neutrino masses are generated through the interactions with the dark sector, resulting in a mass matrix for small $\lambda_5$
\begin{equation}\label{eq:numass}
    (m_\nu)_{\alpha\beta}\approx \frac{\lambda_5 v^2}{32 \pi^2}
\sum_{i=1}^3\frac{y_{i\alpha} y_{i\beta}m_{N_i}}{(m_{R,I}^2-m_{N_i}^2)}
\times \left[1+\frac{m_{N_i}^2}{m_{R,I}^2-m_{N_i}^2}
\ln\left(\frac{m_{N_i}^2}{m_{R,I}^2}\right)\right],
\end{equation}
which is linear in $\lambda_5$. We employ the Casas-Ibarra parameterization \cite{Casas:2001sr} to constrain the Yukawa coupings $y$ using neutrino mass differences and mixings as input \cite{Esteban:2018azc}. We further impose upper bounds on branching ratios (BRs) of lepton flavour violating (LFV) processes, $\mu \to e \gamma$\cite{TheMEG:2016wtm}, $\mu \to 3e$\cite{Bellgardt:1987du}, and $\mu - e, {\rm Ti}$\cite{Dohmen:1993mp}.
The BRs as well as the conversion ratio (CR) are calculated with \textsc{SPheno} 4.0.3 \cite{Porod:2011nf}. The relic density, calculated with \textsc{micrOMEGAs} 5.0.8 \cite{Belanger:2018ccd}, is restricted around its central value of $\Omega h^2=0.12$ \cite{Aghanim:2018eyx}, allowing for a theoretical uncertainty of $\pm 0.02$.

\section{Numerical results}
\label{sec:4}
\begin{figure}[t]
\begin{center}
\includegraphics[width=0.49\textwidth]{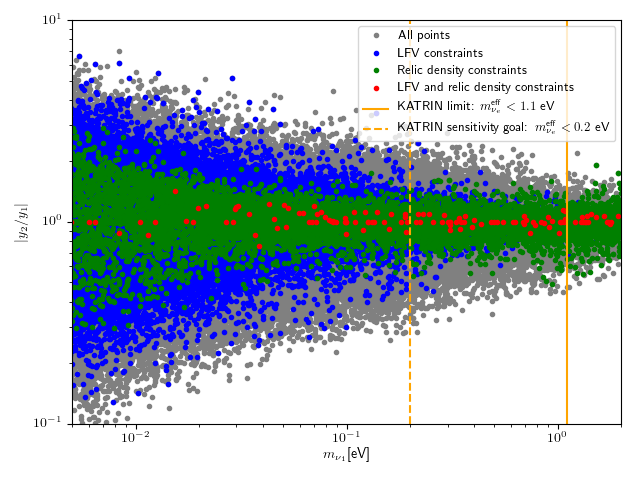}
\includegraphics[width=0.49\textwidth]{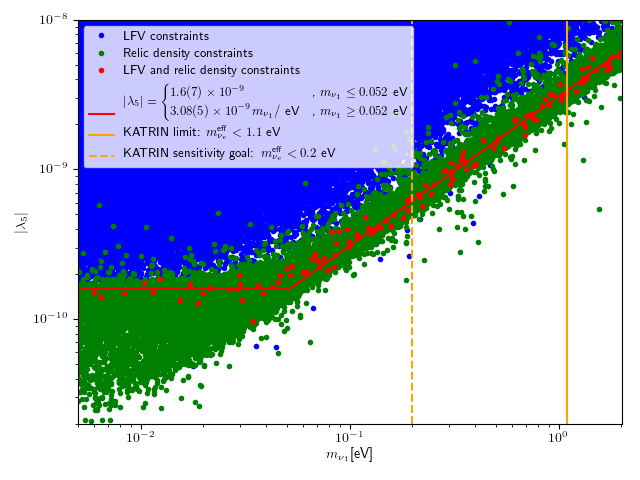}
\vspace{-5mm}
\caption[]{Ratio of Yukawa couplings (left), and the dark sector-Higgs boson coupling $|\lambda_5|$ (right) as a function of lightest
  neutrino mass $m_{\nu_1}$ with mass difference/mixing (grey), LFV (blue), relic density (green) and all constraints (red
  points).}
\label{fig:02}
\end{center}
\end{figure}

The region of an electron neutrino mass of 1.1 eV down to 0.2 eV, which is currently being probed by KATRIN \cite{Aker:2019uuj,Drexlin:2013lha}, approaches the cosmological limit of 0.12 eV\cite{Aghanim:2018eyx}. Here the neutrino mass matrix is dominated by the absolute mass scale. As a result the eigenvalues of the Yukawa matrices $y_\alpha$ become similar in value. Fig.\ \ref{fig:02} (left) shows the ratio $|y_2/y_1|$ (grey points), which varies over the complete $y$ range in the plot for small $m_{\nu_1}$ but the spread decreases to a factor of two at larger masses. Imposing the LFV (blue) and relic density constraints (green) further reduce the spread. Combining all constraints (red) leads to $|y_2/y_1|\sim1$, as well as other ratios of eigenvalues.

\begin{figure}[h]
\begin{center}
\includegraphics[width=0.49\textwidth]{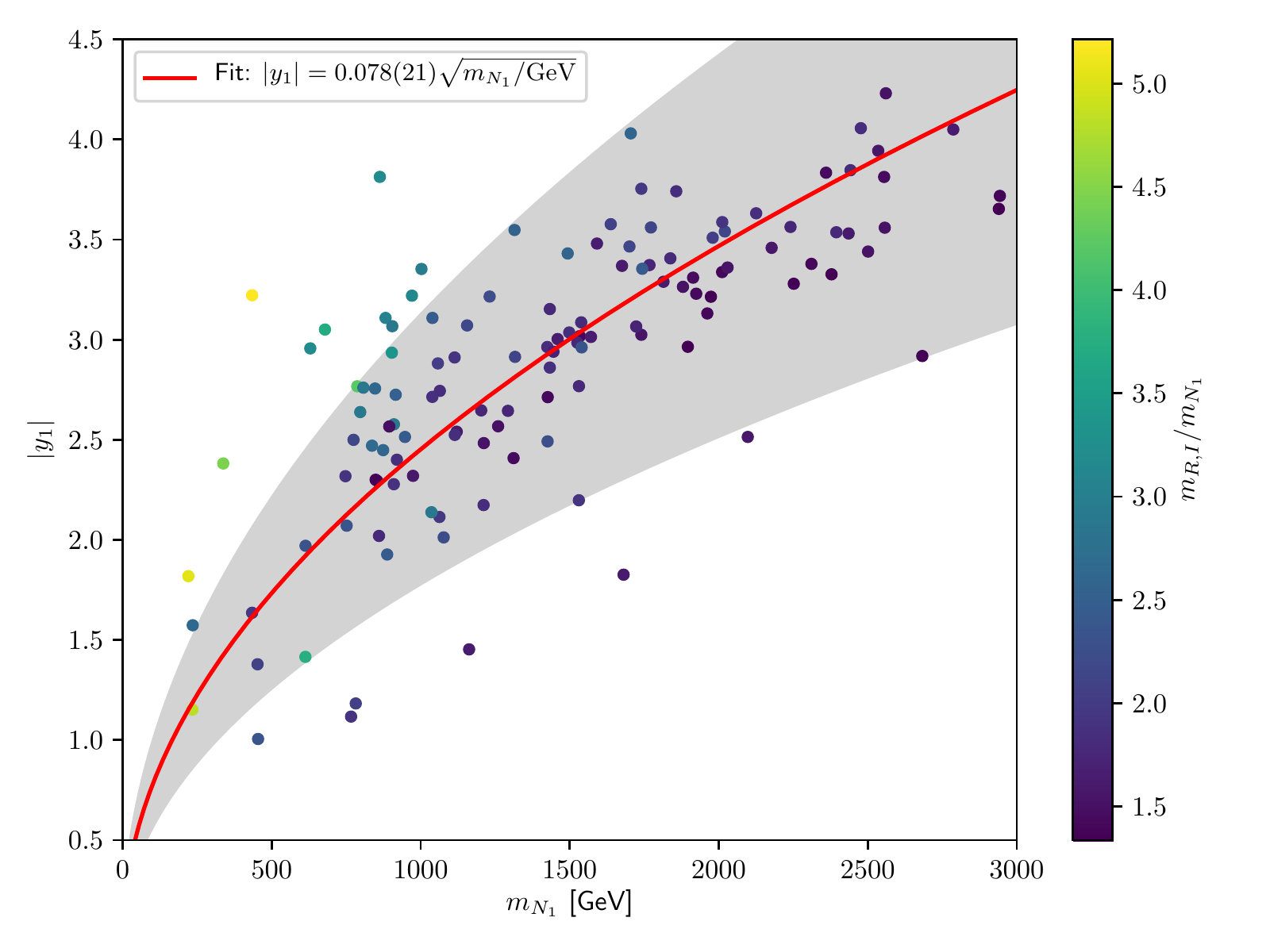}
\vspace{-5mm}
\caption{Yukawa coupling of the lightest neutrino as a function of the DM mass.
  The ratio of the neutral scalar over the DM mass is given on the temperature scale.}
\label{fig:04}
\vspace{-5mm}
\end{center}
\end{figure}

In Eq.\ (\ref{eq:numass}) a linear relation between the neutrino mass matrix and the neutral scalar mass splitting $\lambda_5$ could be observed. This is made explicit in Fig.\ \ref{fig:02} (right) which shows the value of $|\lambda_5|$ as a function of the lightest neutrino mass eigenstate $m_{\nu_1}$. The linear dependence arises after applying the LFV (blue), relic density (green), and both constraints (red). At 90\% C.L.\ its value is (left)
\be
|\lambda_5|=\left\{\begin{array}{ll}
(3.08\pm0.05)\times10^{-9} \ m_{\nu_1}/{\rm eV} & ({\rm NH})\\
(3.11\pm0.06)\times10^{-9} \ m_{\nu_1}/{\rm eV} & ({\rm IH})\end{array}\right., \hspace{3mm}
|\lambda_5|=\left\{\begin{array}{ll}
(1.6\pm0.7)\times10^{-10} & ({\rm NH})\\
(1.7\pm1.5)\times10^{-10} & ({\rm IH})\end{array}\right.
\ee
where the sign is arbitrary. Below a value of $m_{\nu_1}=0.052$ eV the mass of the heaviest neutrino dominates over the mass splittings and $\lambda_5$ does not depend on $m_{\nu_1}$ (right). A KATRIN measurement would therefore result in a prediction for the dark sector-Higgs coupling $\lambda_5$. Combining the results thus far, the value $m_{\nu_1}/|\lambda_5|$ has a fixed value, and the eigenvalues of the Yukawa couplings have approximately equal values. The relic density constrains the ratio of the DM and scalar masses ($m_{R,I}/m_{N_1}\sim1.5$), resulting in a leading term in Eq.\ (\ref{eq:numass}) that is proportional to $|y_1|^2/m_{N_1}$, with $N_2$ and $N_3$ having significantly higher masses. The square root dependence of $|y_1|$ on the DM mass is shown in Fig.\ \ref{fig:04}, with
\be
 |y_1|=\left\{\begin{array}{ll}
 (0.078\pm0.021) \ \sqrt{m_{N_1}/{\rm GeV}} & ({\rm NH})\\
 (0.081\pm0.012) \ \sqrt{m_{N_1}/{\rm GeV}} & ({\rm IH})\end{array}\right. ,
\ee
meaning that a known DM mass results in a prediction of its Yukawa coupling to the SM leptons.

\section{Summary and outlook}
\label{sec:5}
From our results one can show that the parameter space of the fermion DM scotogenic model can almost fully be probed through a combination of a absolute neutrino mass measurement fixing the dark sector-Higgs coupilng, and a DM mass measurement which fixes the Yukawa couplings to the SM leptons. Moreover LFV constraints provide an orthogonal way of probing into the parameter space, severely restricting the remaining parameter space. We checked that our results are valid for both neutrino hierarchies.

As an outlook, our findings can be generalized to different radiative seesaw models, e.g.\ those containing triplet fermions \cite{Ma:2008cu} and/or singlet-doublet scalars \cite{Farzan:2009ji}, where the generation of neutrino masses results in expressions similar to Eq.\ (\ref{eq:numass}).

\vspace{-2mm}
\section*{Acknowledgements}
\vspace{-4mm}
  This work has been supported by the DFG through the Research Training Network 2149
  ``Strong and weak interactions - from hadrons to dark matter''.
%


\end{document}